\documentclass[%
superscriptaddress,
preprint,
 amsmath,amssymb,
 aps,
]{revtex4-2}

\usepackage{graphicx}
\usepackage{dcolumn}

\usepackage{verbatim}
\usepackage{comment}
\usepackage{subcaption}

\usepackage[normalem]{ulem}
\usepackage{xcolor}

\newcommand{\mbf}[1]{\mathbf{#1}}

\newcommand{\varv}{v} 
\newcommand{\reals}{\mathbb{R}} 

\newcommand{\tL}{\mathbf{L}}
\newcommand{\tE}{\mathbf{E}}

\newcommand{\tT}{\mathbf{T}}

\newcommand{\trans}{^\mathrm{T}}

  \newcommand{\bv}{\mbf{\varv}}
  \newcommand{\bx}{\mbf{x}}

  \newcommand{\bL}{\mbf{L}}

  \newcommand{\bff}{\mbf{f}}
  
  \newcommand{\bk}{\mbf{k}}

  \newcommand{\grad}{\boldsymbol{\nabla}} 
  \newcommand{\lap}{\nabla^2} 
  \DeclareMathOperator{\tr}{Tr} 
  \newcommand{\divg}{\grad\cdot} 

  \newcommand{\Wi}{\mathrm{Wi}}
  \newcommand{\Tr}{\mathrm{Tr}}
  
  \newcommand{\bsigma}{\boldsymbol{\sigma}}
  
\newcommand{\bdelta}{\boldsymbol{\delta}}

\newcommand{\bzero}{\mbf{0}}

\newcommand{\balpha}{\boldsymbol{\alpha}}

\newlength{\figwidth}
\newlength{\SCwidth}
\def\XXint#1#2#3{{\setbox0=\hbox{$#1{#2#3}{\int}$}
     \vcenter{\hbox{$#2#3$}}\kern-.5\wd0}}

\begin{document}

\title{Unveiling the stress field in a flowing complex fluid with velocity and microstructure imaging}


\author{Ashinth Mangesh}
\affiliation{Department of Chemical and Biological Engineering,
University of Wisconsin-Madison
}%
\author{Manish Kumar}
\affiliation{Department of Chemical and Biological Engineering,
University of Wisconsin-Madison
}%
\affiliation{Department of Chemical Engineering, Indian Institute of Science, Bangalore 560012, India
}%
\author{Matthew E. Helgeson}
\affiliation{Department of Chemical Engineering, University of California Santa Barbara, Santa Barbara, CA 93106-5080
}%
\author{Michael D. Graham}\email{Corresponding author. Phone: 608-265-3780, email: mdgraham@wisc.edu}%
\affiliation{Department of Chemical and Biological Engineering,
University of Wisconsin-Madison
}%

\date{\today}

\begin{abstract}

The stress in a flowing complex fluid such as a solution of polymers, surfactants and/or colloidal particles is a fundamental dynamic property of the material. It reflects the evolution of the fluid microstructure under deformation and determines the motion of the fluid. While spatially resolved measurements of flow (via velocimetry) and microstructure (e.g.~via scattering) can be made, neither of these individually can uniquely determine the stress. In particular, due to the degenerate nature of the relationship between the stress tensor and the net force exerted by the microstructure on the fluid, the stress tensor is hidden from direct observation.  We show that integrating velocity and microstructure measurements with principles of continuum mechanics can break this degeneracy, unveiling the stress tensor. The approach has two stages. In the first, velocity information is used along with the momentum and mass conservation equations to infer the force density exerted by the microstructure on the fluid using a data assimilation framework. In the second, the equation relating force density to stress is solved. This equation has a nontrivial nullspace, introducing a homogeneous solution whose structure must be found to uniquely determine the stress. We do so by using microstructural data and a general model between microstructure and stress, the parameters of which are determined, along with the nullspace solution, by solving a least-squares problem.  The approach is illustrated using synthetic velocity and microstructural data from simulation of a flowing polymer solution, demonstrating its validity and robustness.   

\end{abstract}

\maketitle



\noindent \textbf{Category}: Physical Sciences/Applied Physical Sciences
~ 

\noindent \textbf{Significance Statement: Many fluids in nature and technology contain micro- or nanoscale components, such as dissolved macromolecules, that make their flow behavior highly complex. Examples include biological fluids such as blood and mucus, solutions of electrically conducting polymers for flexible electronics, and polypeptide solutions for pharmaceutical formulations. The relationship between stress, flow, and microstructure in a complex fluid is fundamental to understanding these fluids and using them effectively in applications, but direct  measurement of stress in general flows is not possible. We show that integrating velocity and microstructure measurements with physical and mathematical principles can reveal this otherwise hidden information, yielding a ``stress imaging" method that will shed light on long-standing scientific questions and provide a concrete tool for optimizing industrial processes.}

~

\noindent The relationship between stress, microstructure and flow is a fundamentally important dynamic physical property of a complex fluid. It is also central to many applications, playing a key role in efficient materials processing as well as in the performance of the final product. 
 Classical rheometry experiments impose only very simple deformations on materials and thus do not expose them to the complex time-dependent deformations that they experience during most flow processes occurring in nature or in engineering applications. 
 Furthermore, neither molecular theory nor semi-empirical constitutive modeling are sufficient to capture the complex evolution of microstructure and stress that occurs in many soft materials, including block copolymers, surfactant solutions, and conducting polymers, among others. Finally, it is not possible to directly measure stress within a material in a general complex flow.
Therefore, inferring the stress in a flowing complex fluid is a long-standing, important, and unsolved problem in soft-matter physics. In general, neither velocimetry nor microstructural measurements alone can provide the stress: it is hidden from direct observation.  The present work demonstrates that, for planar flows, velocimetry and microstructural measurements can be combined with conservation equations and the mathematical structure of the stress tensor to learn the stress-microstructure relationship and thereby reveal the stress field in the fluid.  

Developments in machine learning have led to new and promising advances in constitutive modeling for viscoelastic fluids e.g.~\cite{lennon2023scientific,Mahmoudabadbozchelou:2021bf,Mangal202510112280000908,Sato202510112280000872}. However, these methods have still only been applied to classical rheometric data, and do not capture complex microstructural information beyond the stress tensor. More relevant to the present work, many recent studies have addressed the issue of how to extract constitutive behavior from nontrivial flows. Kumar, Guasto, and Ardekani \cite{kumar2023lagrangian} observed that for solutions of long-chain polymers, the Lagrangian stretching field, which can be obtained from velocimetry, yields substantial information about the polymer stretching and stress. While useful, this approach does not give quantitative estimates of the stress tensor or constitutive model, and is primarily relevant for dilute polymer solutions, for which material line stretching and polymer stress are known to be closely related \cite{Graham.2018.https://doi.org/10.1017/9781139175876}. Thakur, Raissi, and Ardekani \cite{Thakur2024101016jjnnfm2024105265} developed an application of the physics-informed neural network (PINN) approach \cite{Karniadakis2021101038s42254-021-00314-5} in which, given a flow field for a viscoelastic liquid, the parameters of a given constitutive equation that best fit the field are found. In PINNs, the governing equations are incorporated as penalty terms in a loss function that is minimized to fit the model to the data \cite{Karniadakis2021101038s42254-021-00314-5}, with two-dimensional (2D) stenosis and cross-slot flows as the test cases. Sunol et al.~\cite{Sunol20251048550arxiv251024673} addressed a similar class of problems using the JAX framework for differentiable simulation \cite{jax2018github} to determine the gradients in the solution to a flow problem with respect to parameters in a given constitutive model. These gradients are then used with a gradient descent algorithm to optimize similarity to the observed velocity field. This framework was applied to the problem of learning parameters in classical viscoelastic constitutive equations, as well as learning a general inelastic constitutive equation using a tensor basis neural network  \cite{lennon2023scientific,Ling:2016fz}.   Sankaran et al. \cite{Sankaran2026101016jcompfluid2026107037} generalized the PINN approach to a data set from numerical simulation of capillary breakup extensional rheometry (CABER), with or without augmentation with Poiseuille flow simulation data, to determine parameters in either a Newtonian or viscoelastic Phan-Thien-Tanner model. Here the PINN implementation is complex because the flow domain changes with time.  Under certain assumptions, the GENERIC framework for nonequilibrium thermodynamics \cite{OttingerBeyond} can be used to determine a general evolution equation for the conformation tensor and stress in a flowing complex fluid. A key aspect of this equation is the expression for a conformation-dependent entropy. Nieto Simavilla et al.~\cite{Simavilla2025101017jfm202510325} used a PINN approach to determine, as a neural network, this entropy function from flow simulation data where the conformation field was given. 

All of the works described above implicitly assume that the stress tensor or a closely related conformation tensor obeys a closed evolution equation whose form is specified, and used either stress measurements (in the case of rheological experiments) or velocimetry (in the nontrivial flow cases) to determine the parameters of that evolution equation. 
However, even for equations based on simple micromechanical models such as a finitely extensible dumbbell or rigid rod, closure approximations are necessary to find a self-contained evolution equation for the stress \cite{Bird:1987vu,Graham.2018.https://doi.org/10.1017/9781139175876}, and more generally, there is a complex relationship between stress, microstructure, and flow that closed form constitutive models do not capture. 
Young et al.~\cite{Young2023101007s00397-023-01412-0} have described an approach to data-driven determination of microstructural evolution in flow, building on the development of a novel experimental design combining velocimetry and microstructural measurements \cite{Corona2022101103physrevmaterials6045603}. In this experiment, a fluidic four-roll mill  (FFoRM) \cite{Lee:2007fn} compatible with scanning small-angle X-ray scattering (SAXS) measurements is used to collect velocimetry data along with spatially resolved scattering data obtained by scanning the X-ray beam across the flow domain. A key observation from this work was that by examining scattering patterns along streamlines in this ``FFoRM-sSAXS" device, the Lagrangian transient evolution of the microstructure can be observed.  Based on this observation, the authors developed a method, ``Scattering informed microstructure prediction during Lagrangian evolution" (SIMPLE) that uses FFoRM-sSAXS data, along with dimension reduction with an autoencoder \cite{Hinton:2006bg} and the neural ordinary differential equation (NODE) method \cite{chen2018neural}, to find an efficient reduced order and frame indifferent model for a test fluid's microstructural evolution in diverse complex flows.

In general, observations of microstructure, such as the scattering techniques applied in SIMPLE, do not reveal the stress tensor. For some complex fluids, however, including polymer melts and concentrated surfactant solutions near equilibrium, there is is an isotropic linear relationship between the stress tensor and the optical anisotropy or index of refraction tensor known as the \emph{stress-optic rule} \cite{Larson1988va,Fuller1995101093oso97801950971840010001}. Here the stress can be inferred from birefringence images, but only very near equilibrium, and only if the stress-optic coefficient is known. 

Another potential source of spatially resolved microstructural information is fluorescence microscopy of tracer molecules in flow \cite{Smith1999101126science28354081724}. Thorens et al. \cite{Thorens2026} show an initial step in this direction using DNA solutions in a microfluidic cross-slot device. Similar to the work of Young et al. \cite{Young2023101007s00397-023-01412-0}, they study the Lagrangian evolution of the polymer conformations along streamlines in the flow. 

The present work introduces a method to combine spatially resolved velocimetry and microstructural information (whether from scattering, microscopy, or otherwise) in a 2D flow of a complex fluid -- as is available in the FFoRM-sSAXS device -- to simultaneously estimate (1) the stress tensor in the flowing material and (2) the relationship between the stress and the microstructure. We call the approach ``STRess Imaging with Velocimetry and microStructure", or STRIVES. The key advance underlying STRIVES is the resolution of the degenerate nature of the relationship between the stress tensor and the net force exerted by microstructure on the fluid. In the following, we first describe STRIVES, then provide proof of principle for its validity, including robustness in the presence of noise, using simulation data from a well-studied model problem. We conclude by describing further directions and applications.

\section{General formulation}

\subsection{Setting}
\newcommand{\bmu}{\boldsymbol{\mu}}

We consider 2D planar flows, represented using Cartesian coordinates $(x,y)$ in a domain $A$. In this plane we have spatially resolved measurements of velocity $\bv_m(\bx)\in\reals^2$ and microstructure $\bmu(\bx)\in\reals^{N_\mu}$ where $N_\mu$ is the number of microstructural quantities measured at each position. What these quantities are depends on the nature of the microstructural data. For example, this data could be local scattering patterns measured in a FFoRM-sSAXS, as in \cite{Corona2022101103physrevmaterials6045603} and \cite{Young2023101007s00397-023-01412-0}, in which case $N_\mu$ is the number of pixels in the scattering pattern. We will assume that the given microstructural information is sufficient in principle to determine the microstructural contribution to the stress.  For a complex fluid (containing polymers, surfactants, colloidal particles,\ldots) in the absence of inertia, the nondimensional 2D continuum momentum and continuity equations, at scales large compared to microstructural scales, are 
\begin{equation}
    \bzero=-\grad p+\beta\lap\bv+\gamma\bff_\mu+\bff_e, \qquad    0=\divg\bv.
\label{eq:Cauchy}
\end{equation}
Here $p$ is the pressure, $\bff_e$ a given external forcing, and 
\begin{equation}
    \bff_\mu=\divg\bsigma_\mu
    \label{eq:stressdiv}
\end{equation} is the force density arising from the $2\times 2$ symmetric microstructural stress tensor $\bsigma_\mu$, which ultimately is the quantity we wish to determine. (While the $zz$ component of stress can be nonzero in planar flows, it does not enter in the 2D momentum equation, so it does not affect the flow and accordingly cannot be determined from it.) The quantities $\beta$ and $\gamma$ are nondimensional solvent viscosity and microstructural stress scalings that arise from the details of the nondimensionalization. In the original dimensional representation $\beta$ is replaced by the solvent viscosity $\eta_s$, and $\gamma=1$. Using the Helmholtz decomposition, $\bff_\mu$ can be rewritten as the sum of solenoidal and irrotational components
\begin{equation}
    \bff_\mu=\bff_\mu^s-\grad \xi, \quad\textrm{where~}\divg\bff_\mu^s=0.
    \label{eq:stresshelm}
\end{equation}
In an incompressible flow, $\xi$ is indeterminate as it can be absorbed into $p$, which we do from here on. 
 We assume that stress and microstructure are related by some as yet undetermined function: i.e.
\begin{equation}
    \bsigma_\mu=\tT(\bmu).
    \label{eq:stressrel}
\end{equation}
The formulation we consider can be broadened to encompass inertia or unsteadiness. The former would add nonlinearity to the problem, which can be accommodated straightforwardly, although for the fluids and flow geometries of interest, we expect inertia to be negligible. The latter would be very complex to include, leading to time-dependent optimization problems, so our focus here will be on steady flows. The stress may also depend on the local strain rate $\tE$ as well as the microstructure; this dependence can be incorporated, but we do not do so here.  

The STRIVES methodology has two parts: the first is to determine the solenoidal force density $\bff_\mu^s$ from velocity data and Eq.~\eqref{eq:Cauchy}; the second, and more significant, is to determine $\bsigma_\mu$ and $\tT(\mu)$ from $\bff_\mu(\bx)$, $\mu(\bx)$ and Eq.~\eqref{eq:stressdiv}.   We address these in turn. 

\subsection{Data assimilation to determine force density}\label{sec:DAforForceDensity}
\newcommand{\fmus}{\bff_\mu^s}
\newcommand{\pt}{p}
To determine $\fmus$ from measurements $\bv_m$ of the velocity field, a variational data assimilation framework \cite{Asch:2016,Kontogiannis.2022.10.1017/jfm.2022.503} is used here. 
Our aim is to find the force density $\fmus$ that generates a velocity field $\bv$ minimizing the integrated squared error between the measured and predicted velocities: 
\begin{equation}
    J=\frac{1}{2}\langle||\bv-\bv_m||^2\rangle,
\end{equation}
where $\langle\cdot\rangle=\int_A\cdot\;d\bx$, 
subject to the constraint that $\bv$ and $\fmus$ satisfy Eq.~\eqref{eq:Cauchy}. To minimize $J$, we use gradient descent, determining the gradient of $J$ with respect to $\fmus$ using an adjoint-based approach. Alternately, a differentiable flow solver could be used. For the present analysis, we consider boundary conditions on the flow problem such that the Stokes operator is self-adjoint. For example, periodic or no-slip boundary conditions satisfy this condition; the example below is periodic.

Real velocity measurements will have noise that may pollute the computation of $\fmus$. There are many ways to address this problem, including pre-filtering the velocity measurements before computing $\fmus$. Another, which we apply below, is to add a small gradient penalty term to $J$ of the form
\begin{equation}
    \frac{\lambda_n}{2}\langle\grad\fmus:\grad\fmus\rangle.
    \label{eq:fgradient}
\end{equation}

\subsection{Determining stress from force density and microstructure}

Understanding the solution structure of Eq.~\eqref{eq:stressdiv} is one of the keys to determining $\bsigma_\mu$. We first observe that the divergence operator is singular; the equation $\divg\bsigma_\mu=\bzero$ has a nullspace of nontrivial solutions 
\begin{equation}
    \bsigma_\mu^n=\begin{bmatrix}
\frac{\partial^2 \chi}{\partial y^2} & -\frac{\partial^2\chi}{\partial x\partial y}\\-\frac{\partial^2\chi}{\partial x\partial y} & \frac{\partial^2 \chi}{\partial x^2},
\end{bmatrix}\label{eq:stressnull}
\end{equation}
where $\chi(\bx)$ is an as yet arbitrary scalar field. Without further information, the stress tensor is thus indeterminate, as any choice of $\chi$ can solve Eq.~\eqref{eq:stressdiv}. In classical elasticity, with $\bsigma_\mu$ replaced by the total stress tensor, this quantity is the \emph{Airy stress function} \cite{SaddElasticity2005}. A central insight of the present work is the recognition that information about $\chi(\bx)$ is encoded in the microstructure of the fluid.

In addition to this homogeneous solution, Eq.~\eqref{eq:stressdiv} has particular solutions arising from each piece of the Helmholtz decomposition of $\bff_\mu$:
\begin{align}
    \bsigma_\mu^s~&\textrm{~solves~}\divg\bsigma_\mu^s=\bff_\mu^s,\label{eq:particular1}\\
    \bsigma_\mu^i=-\xi\bdelta&\textrm{~solves~}\divg\bsigma_\mu^i=-\grad\xi.
\end{align}
Thus the general solution to Eq.~\eqref{eq:stressdiv} is given by
\begin{equation}
    \bsigma_\mu=\sigma_\mu^s-\xi\bdelta+\begin{bmatrix}
\frac{\partial^2 \chi}{\partial y^2} & -\frac{\partial^2\chi}{\partial x\partial y}\\-\frac{\partial^2\chi}{\partial x\partial y} & \frac{\partial^2 \chi}{\partial x^2}
\end{bmatrix}.
\end{equation}
To eliminate $\xi$, which is inaccessible, we rewrite this in terms of the microstructural contribution to the shear stress $\tau=\sigma_{\mu ,yx}$ and normal stress difference $\Sigma=\sigma_{\mu ,xx}-\sigma_{\mu ,yy}$:
\begin{equation}
    \begin{split}
    \tau&=\sigma_{\mu ,yx}^s -\frac{\partial^2\chi}{\partial x\partial y},\\
\Sigma&=\sigma_{\mu ,xx}^s-\sigma_{\mu ,yy}^s+\left(\frac{\partial^2}{\partial y^2}-\frac{\partial^2}{\partial x^2}\right)\chi.
\end{split}
\label{eq:tausigma}
\end{equation}
From the velocimetry data alone, we can determine $\bff_\mu^s(\bx)$ and thus, by solving Eq.~\eqref{eq:particular1}, $\bsigma_\mu^s(\bx)$. The function $\chi(\bx)$ remains indeterminate. 

However, in the present context we also have microstructural information, and the stress tensor, including $\chi(\bx)$ has to be consistent with this. We now describe a framework for determining $\chi(\bx)$, and thus $\tau$ and $\Sigma$, as well as the relationship between stress and microstructure.  In the present work we will do so in Fourier space for application to periodic geometries. Future studies will develop a finite-element-based computational framework for general geometries.

In a square periodic domain with sides of length $2\pi$, Eq.~\eqref{eq:stressdiv}, transformed into Fourier space, becomes
\begin{equation}
  i\bk\cdot\bsigma_\mu(\bk)=\bff_\mu(\bk),
    \label{eq:stressdivk}
\end{equation}
where $\bk=[k_x,k_y]\trans$. (We indicate whether quantities are in physical or Fourier space representation by their argument, $\bx$ or $\bk$, respectively.)  In Fourier space, the divergence-free condition on $\bff_\mu^s$ becomes 
\begin{equation}
    i\bk\cdot\bff_\mu^s(\bk)=0, \textrm{or}~\bff_\mu^s(\bk)\perp \bk,
\end{equation} so, defining $\bk^\perp=[-k_y,k_x]\trans$,  $\bff_\mu^s(\bk)$ can be written
\begin{equation}
    \bff_\mu^s(\bk)=\alpha(\bk)\bk^\perp, \quad \alpha(\bk)=\frac{\bff_\mu^s(\bk)\cdot\bk^\perp}{||\bk||^2}.
\end{equation}
With this forcing, we can solve Eq.~\eqref{eq:stressdivk} to yield, for $k_x\neq 0$:
\begin{equation}
    \bsigma_\mu^s(\bk)=-i\alpha(\bk)\begin{bmatrix}
    -\frac{2k_y}{k_x} & 1 \\
    1 & 0
    \end{bmatrix},
\end{equation}
and for $k_x=0$: 
\begin{equation}
    \bsigma_\mu^s(0,k_y)=-i\alpha(0,k_y)\begin{bmatrix}
    0 & 1 \\
    1 & 0
    \end{bmatrix}. 
\end{equation}
With these solutions, we can write \eqref{eq:tausigma} as 
\begin{equation}
\begin{split}
\tau(\bk)&=\sigma_{\mu ,yx}^s(\bk) -k_x k_y\chi(\bk),\\
\Sigma(\bk)&=\sigma_{\mu ,xx}^s(\bk) - \sigma_{\mu ,yy}^s(\bk)+\left(k_y^2-k_x^2\right)\chi(\bk).
\end{split}
\label{eq:stresssolutionk}
\end{equation}

At $\bk=\bzero$, $\fmus$ is zero, because it arises from force \emph{dipoles} exerted by the microstructure.

As noted above, we do not know a priori a model relating stress and microstructure, but we assume that the microstructural information available is sufficient in principle to determine the stress. Let us further assume for the present a {linear} model
\begin{equation}
\begin{bmatrix}
    \tau\\ \Sigma
\end{bmatrix}
=\begin{bmatrix}
    \bL_\tau^T \\ \bL_\Sigma^T
\end{bmatrix}\bmu\equiv\tL\bmu
    \label{eq:stressmodel}
\end{equation}
where $\bL_\tau\in \reals^{N_\mu}$ and $\bL_\Sigma\in \reals^{N_\mu}$ are vectors of  parameters to be determined below. The restriction to a linear model is not necessary, but simplifies the formulation and can still represent a broad class of functional forms if we expand $\bmu$ to include nonlinear functions of the measured microstructural quantities, as we do below.  Combining Eq. \eqref{eq:stressmodel} with Eq.~\eqref{eq:stresssolutionk} and rearranging yields
\begin{equation}
\begin{split}
\bL_\tau^T\bmu+k_x k_y\chi(\bk)&=\sigma_{\mu ,yx}^s(\bk) ,\\
\bL_\Sigma^T\bmu+\left(k_x^2-k_y^2\right)\chi(\bk)&=\sigma_{\mu ,xx}^s(\bk) - \sigma_{\mu ,yy}^s(\bk).
\end{split}
\label{eq:stressLmodel}
\end{equation}
 For $N_k$ nonzero wave vectors, this is $2N_k$ equations. Unknowns are the $2N_\mu$ elements of $\bL$ and the value of  $\chi(\bk)$ at each wave vector, for a total of $2N_\mu+N_k$.  In general, because the model is not exact, and because of errors and noise in the data for $\bv$ and $\bmu$, these equations cannot be solved exactly. Nevertheless, as long as $N_k>2N_\mu$, we can pose a linear least squares regression problem based on Eq.~\eqref{eq:stressLmodel} to self-consistently estimate $\bL_\tau$, $\bL_\Sigma$, and $\chi(\bx)$. Furthermore, data from several experiments can be combined into a single least-squares problem -- in that case $N_k$ is the total number of wavenumbers/mesh points from all of the experiments.  

A nonlinear model (e.g.~a neural network) could be used in place of Eq.~\eqref{eq:stressmodel}, resulting in a nonlinear regression problem. (This would need to be discretized in physical space rather than Fourier space.) If the microstructural variables contain redundant information with respect to the stress, then a regularized regression scheme (weight decay, ridge regression, LASSO)\cite{Strang2019} will be appropriate. Regardless of the details of the regression problem, the basic structure introduced here, i.e.~a parameterized stress-microstructure relation and simultaneous solution for the model parameters and nullspace function $\chi(\bx)$, is the enabling feature of our approach, which we call ``STRess Imaging with VElocimetry and microStructure" (STRIVES). 

    From the data, we never have stress, only the divergence-free part of polymer force density. (This is assuming we do not have simultaneous rheology and microstructural data such as that obtained from rheo-SAXS or rheo-SANS. If that is available, it can be used to validate the model in viscometric flows and/or augment the data set used to train the model.) The model prediction for the deviatoric part of $\bsigma_\mu$ is $[\Sigma, \tau; \tau, -\Sigma]$, where we have found $\Sigma$ and $\tau$ from the above regression problem.  Correspondingly, we can compute the predicted force density $\tilde{\bff}_\mu$ and its solenoidal part $\tilde{\bff}_\mu^s$. In Fourier space, with $\hat{\bk}=\bk/||\bk||$,
    \begin{equation}
        \tilde{\bff}_\mu^s=\left(\bdelta-\hat{\bk}\hat{\bk}\right)i\bk\cdot\begin{bmatrix}
            \Sigma & \tau \\ \tau & -\Sigma
        \end{bmatrix},
    \end{equation}
    from which the prediction error $\tilde{\bff}_\mu^s-\bff_\mu^s$ can be computed.

\subsection{Model problem}
We illustrate application of the STRIVES framework with a model cellular flow of a dilute polymer solution that has been widely studied because of the simple geometry and forcing, and the strongly localized elastic stresses and elasticity-driven flow instabilities that can arise \cite{Thomases:2007fl}. The domain is a square of side $L=2\pi$. An external forcing $F\bff_e$, where
\begin{equation}
    \bff_e = \begin{bmatrix}-2\sin(x)\cos(y)\\ 2\cos(x)\sin(y)\end{bmatrix},
\end{equation}
drives the flow of a fluid whose microstructural stress is given by the FENE-P model with relaxation time $\lambda$ \cite{Bird:1987vu}. The zero-rate polymer contribution to the shear viscosity is $\eta_p$, and the total zero-shear rate viscosity $\eta_t=\eta_s+\eta_p$. Defining a characteristic velocity $U=L^2 F/\eta_t$, stress scale $\eta_t U/L$,  time scale $L/U$, and Weissenberg number $\Wi=\lambda U/L$, the  nondimensional momentum equation is Eq.~\eqref{eq:Cauchy} with $\beta=\eta_s/\eta_t$ and $\gamma=(1-\beta)/\Wi$. 
The microstructural stress is given by
\begin{equation}
    \bsigma_\mu = \left(\frac{\boldsymbol{\alpha}}{1-\tr\boldsymbol{\alpha}/b}-\bdelta\right),
\label{eq:fenep_nodim}
\end{equation}
where $\balpha$ is a conformation tensor and $b$ represents the maximum extensibility of the polymeric chains.  
The conformation tensor evolves according to
\begin{equation}
\frac{\partial \boldsymbol{\alpha}}{\partial t}+\bv\cdot \nabla \boldsymbol{\alpha} - \boldsymbol{\alpha}\cdot \nabla \bv-(\boldsymbol{\alpha}\cdot \nabla \bv)^T =-\frac{1}{\Wi}\bsigma_\mu.
\label{eq:conformation}
\end{equation}
This framework has been implemented using a Fourier spectral method with the \emph{Dedalus} software \cite{Burns.2020.10.1103/physrevresearch.2.023068}. Each variable is represented with $64$ Fourier basis functions in each direction, and we find steady states by time-integration from equilibrium initial conditions. 

\section{Results}
\begin{figure}[htbp]
    \centering
    \begin{subfigure}[b]{0.49\textwidth}
        \centering
        \includegraphics[width=\textwidth]{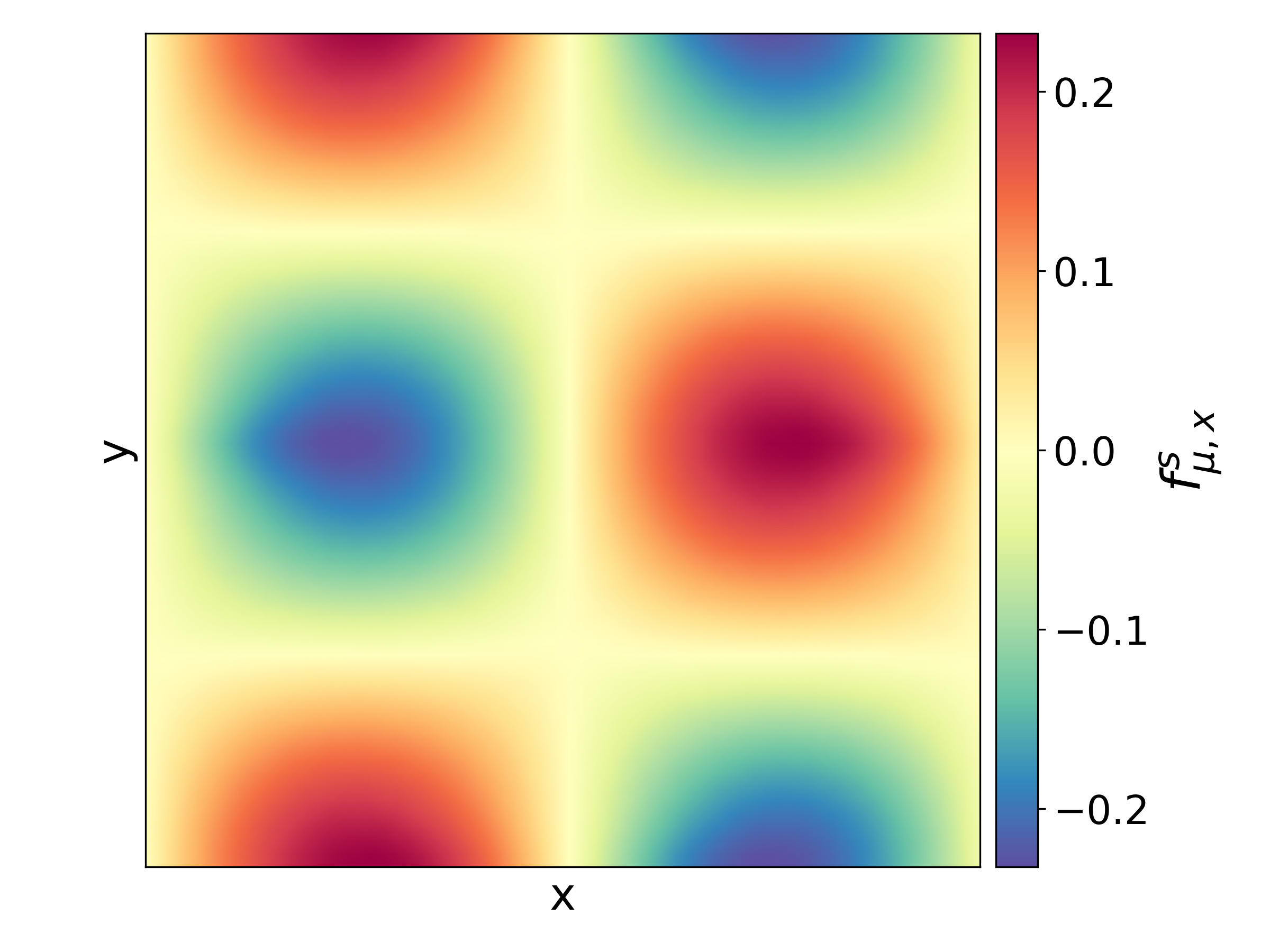}
        \caption{$f_{\mu,x}^s$}
        \label{fig:F_x_adj_Wi03}
    \end{subfigure}
    \hfill 
    \begin{subfigure}[b]{0.49\textwidth}
        \centering
        \includegraphics[width=\textwidth]{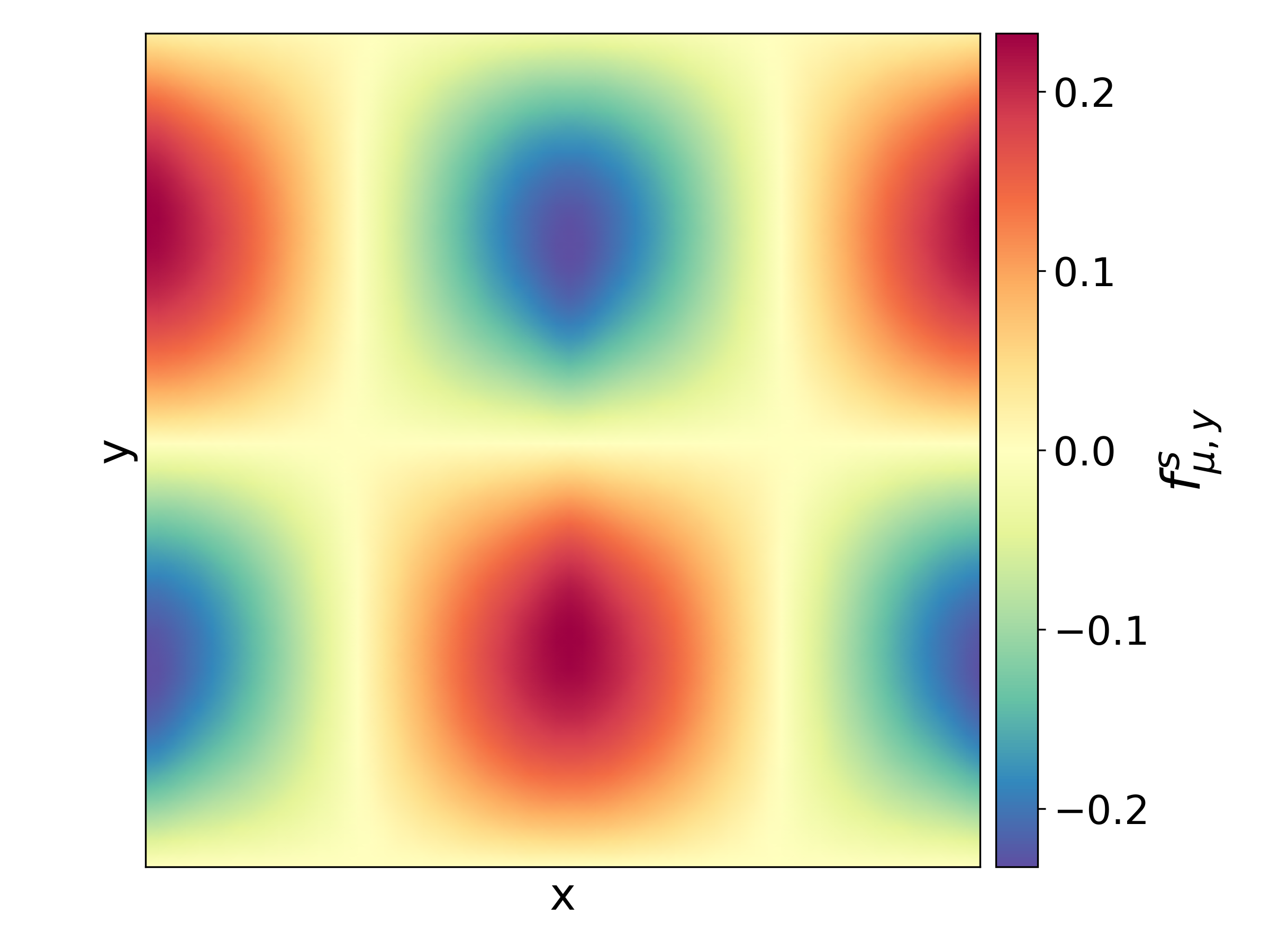}
        \caption{$f_{\mu,y}^s$}
        \label{fig:F_y_adj_Wi03}
    \end{subfigure}
    \hfill 

    \vspace{0.5cm} 
    
        \begin{subfigure}[b]{0.49\textwidth}
        \centering
        \includegraphics[width=\textwidth]{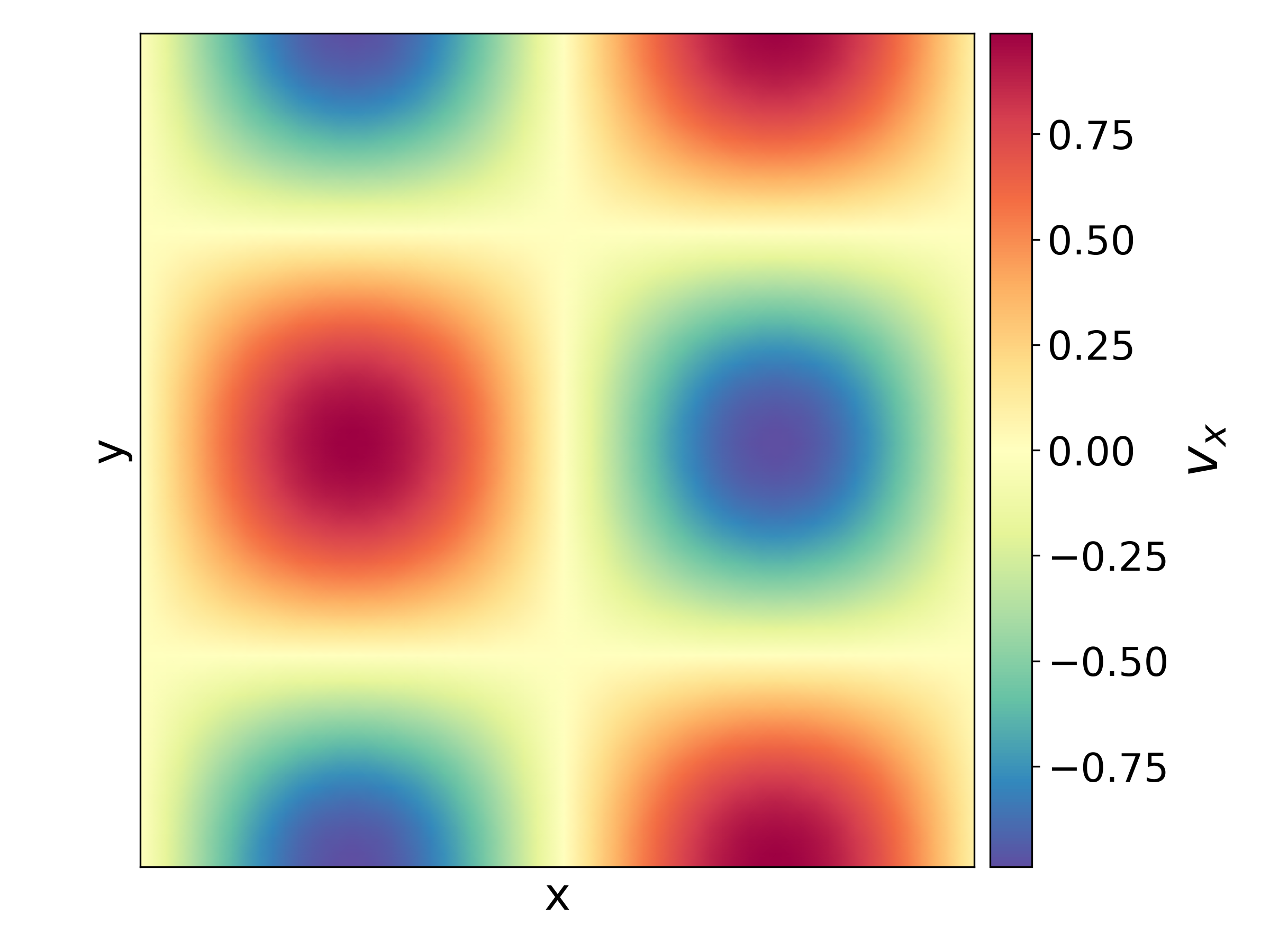}
        \caption{$v_{x}$}
        \label{fig:U_x_adj_Wi03}
    \end{subfigure}
    \hfill 
    \begin{subfigure}[b]{0.49\textwidth}
        \centering
        \includegraphics[width=\textwidth]{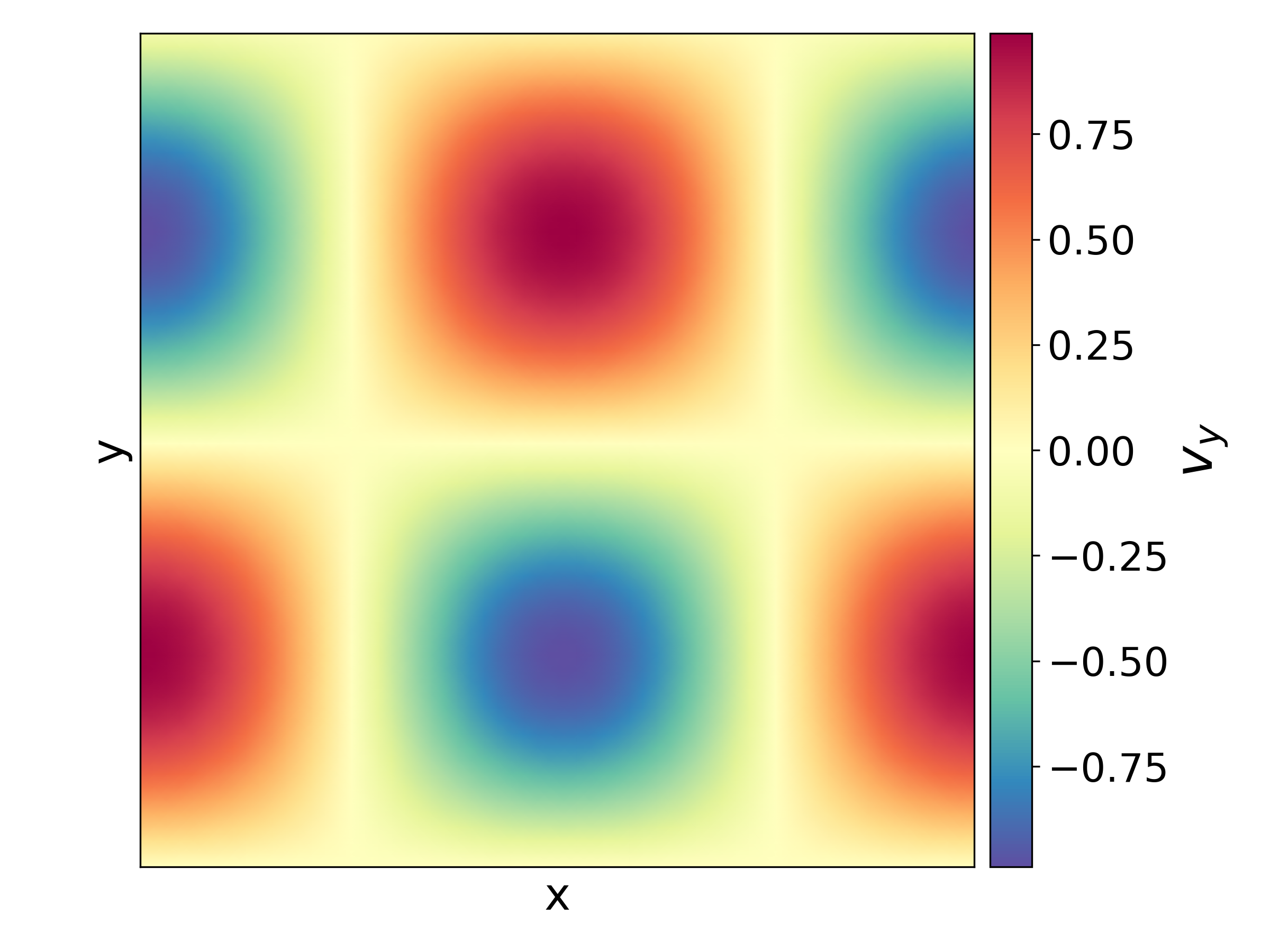}
        \caption{$v_{y}$}
        \label{fig:U_y_adj_Wi03}
    \end{subfigure}
    \hfill
    
    \caption{Solenoidal force density (a-b) and velocity components (c-d) from data assimilation.}
    \label{fig:DAresults1}
\end{figure}

Here we illustrate application of STRIVES to results for the cellular flow problem with $\Wi=0.3, \beta=0.9, b=6400$.  Fig.~\ref{fig:DAresults1} shows the velocity field $\bv$ and solenoidal force density $\fmus$ computed using the methodology of Sec.~\ref{sec:DAforForceDensity}. For all components the maximum absolute error is $\sim 10^{-7}$, indicating the accuracy of this approach. This is expected; the basic approach of Sec.~\ref{sec:DAforForceDensity} is standard and widely used. 

Now we turn to the key findings of this investigation. First, as a simple illustration, we take the vector of observables to be a constant bias and the stress components themselves:
\begin{equation}
    \bmu = \left[ 1 , \sigma_{\mu,xx}, \sigma_{\mu,yx}, \sigma_{\mu,yy} \right]^T.
\end{equation} 
In this case the true value of the model parameter matrix is simply
\begin{equation}
    \tL=\begin{bmatrix}
    \bL_\tau^T \\ \bL_\Sigma^T
\end{bmatrix}=\begin{bmatrix}
    0 & 0 & 1 &0\\ 0 & 1 & 0 & -1  
\end{bmatrix}.\label{eq:exactL}
\end{equation}
Solution of the least-squares problem, Eq.~\eqref{eq:stressLmodel} for the components of $\tL$ yields nearly the exact result:
\begin{equation}
    \tL=\begin{bmatrix}
    \bL_\tau^T \\ \bL_\Sigma^T
\end{bmatrix}=\begin{bmatrix}
    0 & 0 & 1 &0\\ 0 & 1 & 0 & -1  
\end{bmatrix}+\begin{bmatrix}
    -1.1\cdot 10^{-10} & 1.47\cdot10^{-9} & 5.02\cdot 10^{-5} &7.6\cdot 10^{-9}\\ 9\cdot 10^{-9} & -0.56\cdot 10^{-5} & 7.7\cdot 10^{-9} & 0.074\cdot 10^{-5}
\end{bmatrix}\label{eq:LSTRIVES}
\end{equation}


Given that in this simple case,  STRIVES recovers the stress-microstructure relation $\tL$ essentially exactly, it will also reconstruct with high accuracy the fields $\Sigma$ and $\tau$ from the microstructure via Eqs.~\eqref{eq:stressmodel} and \eqref{eq:LSTRIVES}.

\begin{figure}[t]
    \centering
    \begin{subfigure}[b]{0.49\textwidth}
        \centering
        \includegraphics[width=\textwidth]{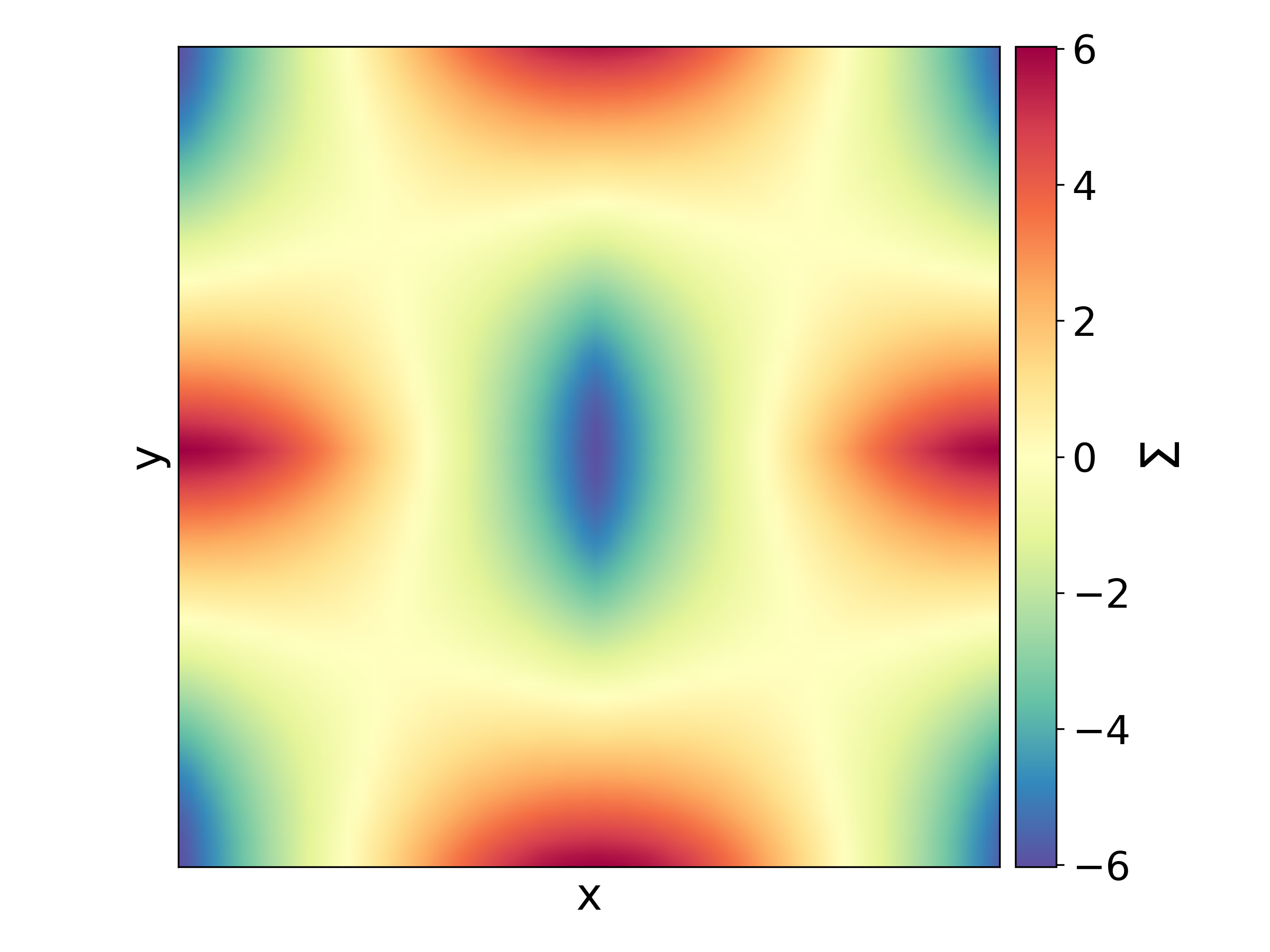}
        \caption{Reconstructed $\Sigma$}
        \label{fig:N1_Model_Wi03}
    \end{subfigure}
    \hfill 
    \begin{subfigure}[b]{0.49\textwidth}
        \centering
        \includegraphics[width=\textwidth]{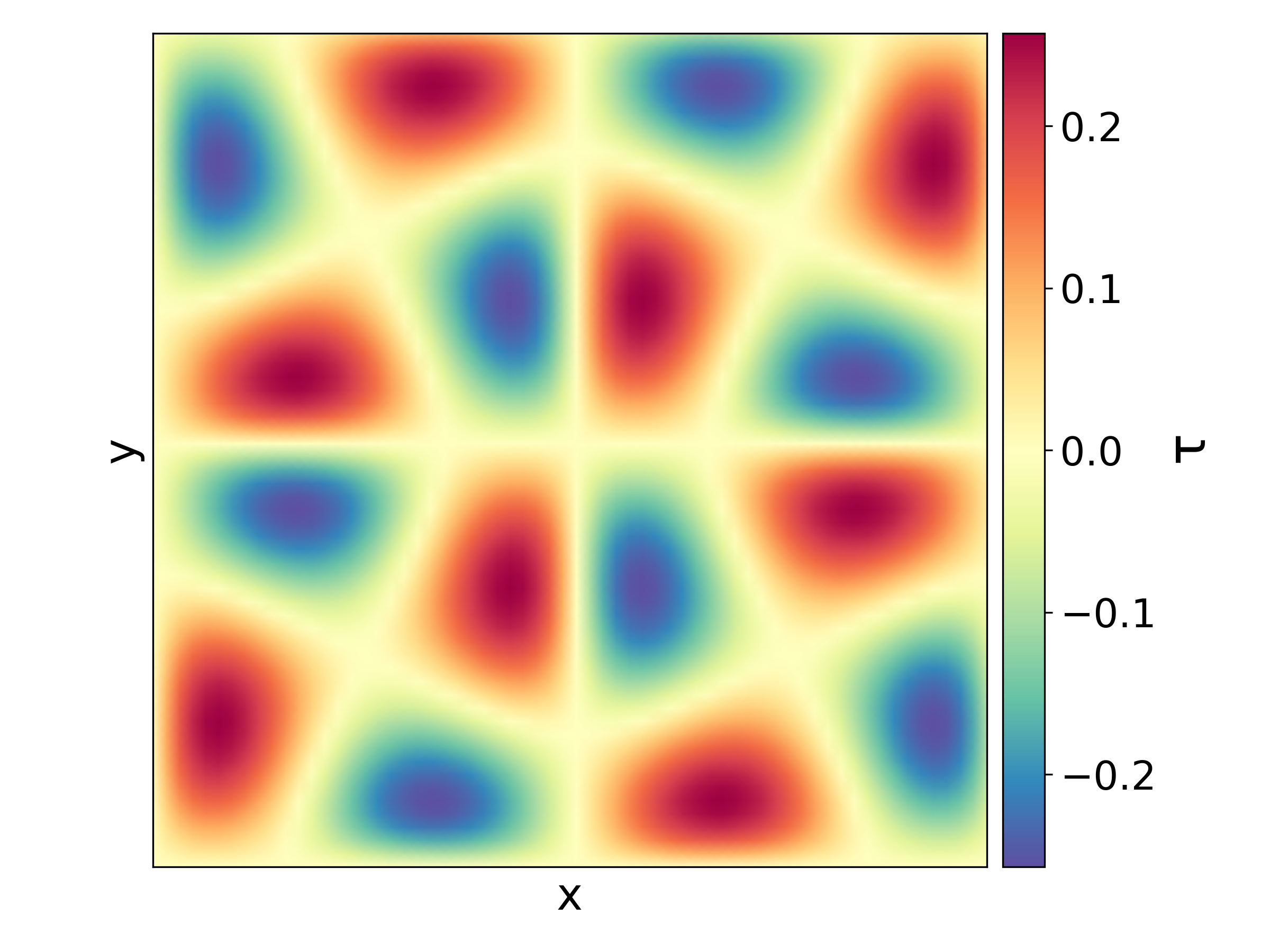}
        \caption{Reconstructed $\tau$}
        \label{fig:shear_Model_Wi03}
    \end{subfigure}
    \caption{ Normal stress difference $\Sigma$ (a) and shear stress $\tau$ (b) reconstructed from Eqs.~\eqref{eq:stressmodel} and \eqref{eq:Lalphaobservables}, i.e., using the observables given in Eq.~\eqref{eq:confobservables}. Absolute errors in the two cases are less than $3\cdot 10^{-4}$ and $6\cdot 10^{-4}$  for $\Sigma$ and $\tau$, respectively. }
    \label{fig:stressreconWi03}
\end{figure}

We now consider a nontrivial microstructural observable. The conformation tensor $\balpha$ is the ensemble-averaged second moment of the polymer end-to-end distance in the FENE-P model. (In a Gaussian approximation, this quantity can be directly estimated from the structure factor.)  The stress is not a linear function of the conformation tensor, so to accurately capture the relation between stress and microstructure, while maintaining the linear model structure described above, we must include nonlinear functions of $\balpha$ in the microstructural variable $\bmu$. Motivated by the Taylor expansion of the Peterlin correction in Eq.~\eqref{eq:fenep_nodim}, $(1-\tr\balpha/b)^{-1}\approx 1+\tr\balpha/b+(\tr\balpha/b)^2+O((\tr\balpha/b)^3)$, we consider the microstructural observable
\begin{equation}
\bmu=\begin{bmatrix}
1& \alpha_{xx}& \alpha_{xy}&\alpha_{yy} & \Tr(\balpha)\alpha_{xx} & \Tr(\balpha)\alpha_{xy}& \tr(\balpha)\alpha_{yy} &
         \tr(\balpha)^2\alpha_{xx} &\tr(\balpha)^2\alpha_{xy} \, \tr(\balpha)^2\alpha_{yy}\end{bmatrix}^T.
         \label{eq:confobservables}
\end{equation}
The STRIVES solution for the stress-microstructure model is now (replacing all values with magnitude $<10^{-8}$ with zeroes)
\begin{equation}
    \tL=\begin{bmatrix}
     0 & 0 & 1.224 & 0 & 0 
        & -0.335 & 0 & 0 & 0.248 & 0 \\
        0 & 1.113 & 0 & -1.113 & -4.37\cdot 10^{-3} 
        & 0 & 4.53\cdot 10^{-3} & 3.14\cdot 10^{-3} & 0 & -3.22\cdot 10^{-3}
        \end{bmatrix}.
        \label{eq:Lalphaobservables}
\end{equation}
The coefficients that are linear in the components of $\balpha$ are largest, with the nonlinear terms providing small corrections to account for the nonlinear relationship between $\bsigma_\mu$ and $\balpha$.
Fig.~\ref{fig:stressreconWi03} shows the reconstructed fields for these quantities; absolute errors are less than $3\cdot 10^{-4}$ and $6\cdot 10^{-4}$  for $\Sigma$ and $\tau$, respectively.

\begin{figure}[t]
    \centering
    \begin{subfigure}[b]{0.49\textwidth}
        \centering
        \includegraphics[width=\textwidth]{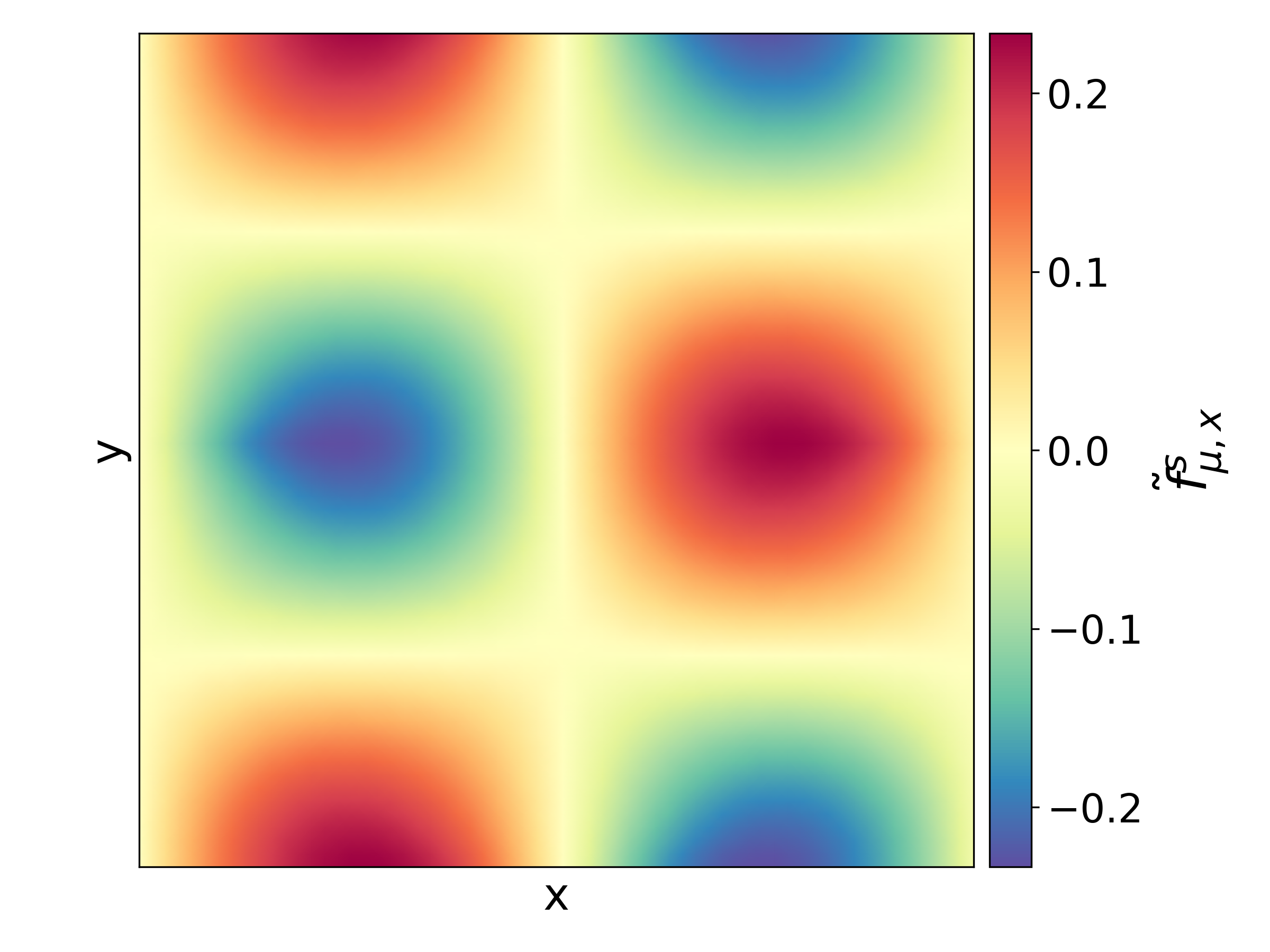}
        \caption{$\tilde{f}_{\mu,x}^s$}
        \label{fig:Fx_perp_model_Wi03}
    \end{subfigure}
    \hfill 
        \begin{subfigure}[b]{0.49\textwidth}
        \centering
        \includegraphics[width=\textwidth]{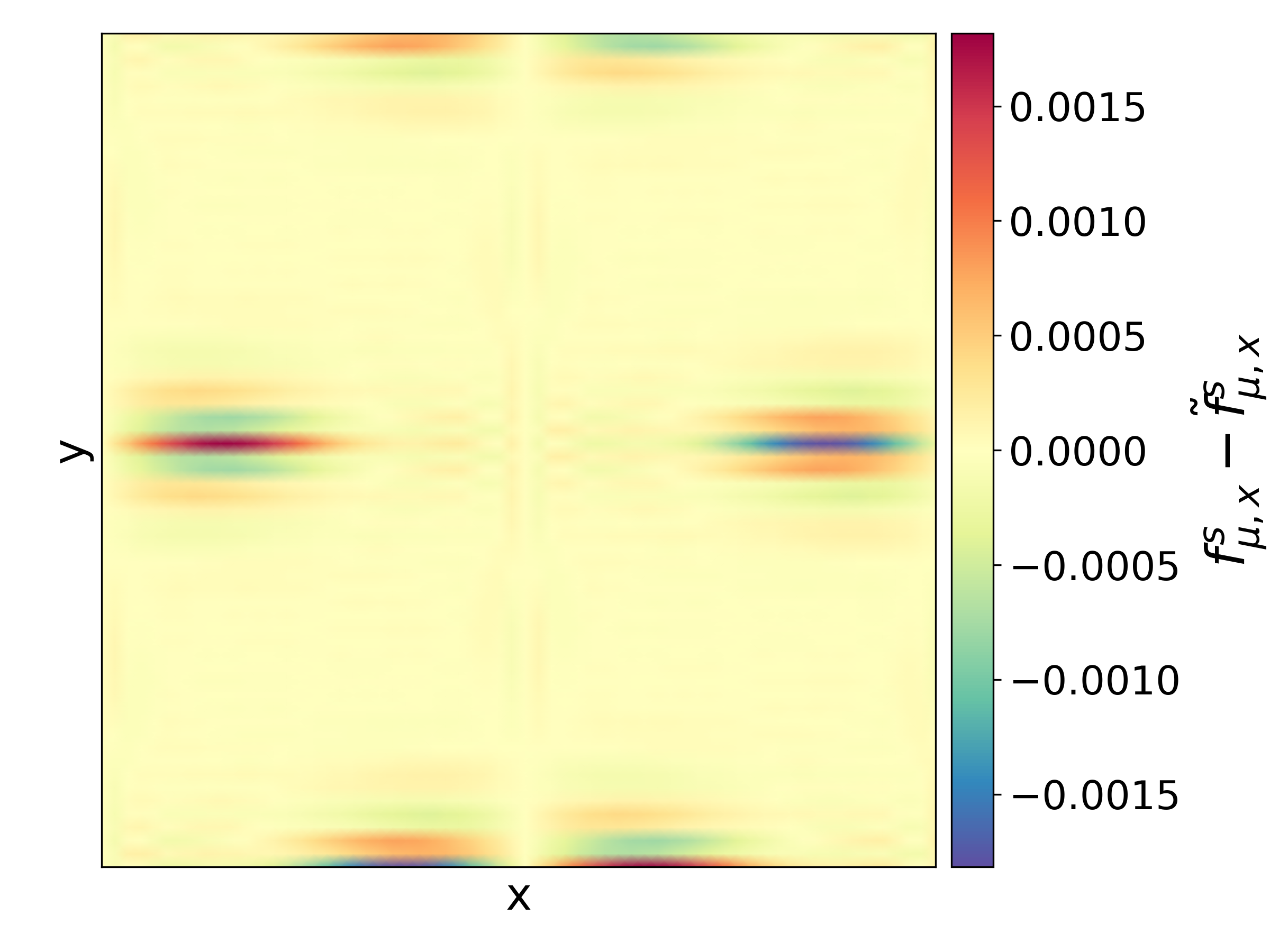}
        \caption{Pointwise difference for $\tilde{f}_{\mu,x}^s$}
        \label{fig:Fx_perp_diff_Wi03}
    \end{subfigure}
    \hfill 

    \caption{(a) STRIVES reconstruction of $x$-component of  polymer force density $f_{\mu}^s$, using the stress reconstructed from conformation tensor observables (Eq.\eqref{eq:confobservables}). (b) Difference from true value.}
    \label{fig:fmudiff}
\end{figure}

\begin{figure}
    \centering
        \begin{subfigure}[b]{0.49\textwidth}
        \centering
        \includegraphics[width=\textwidth]{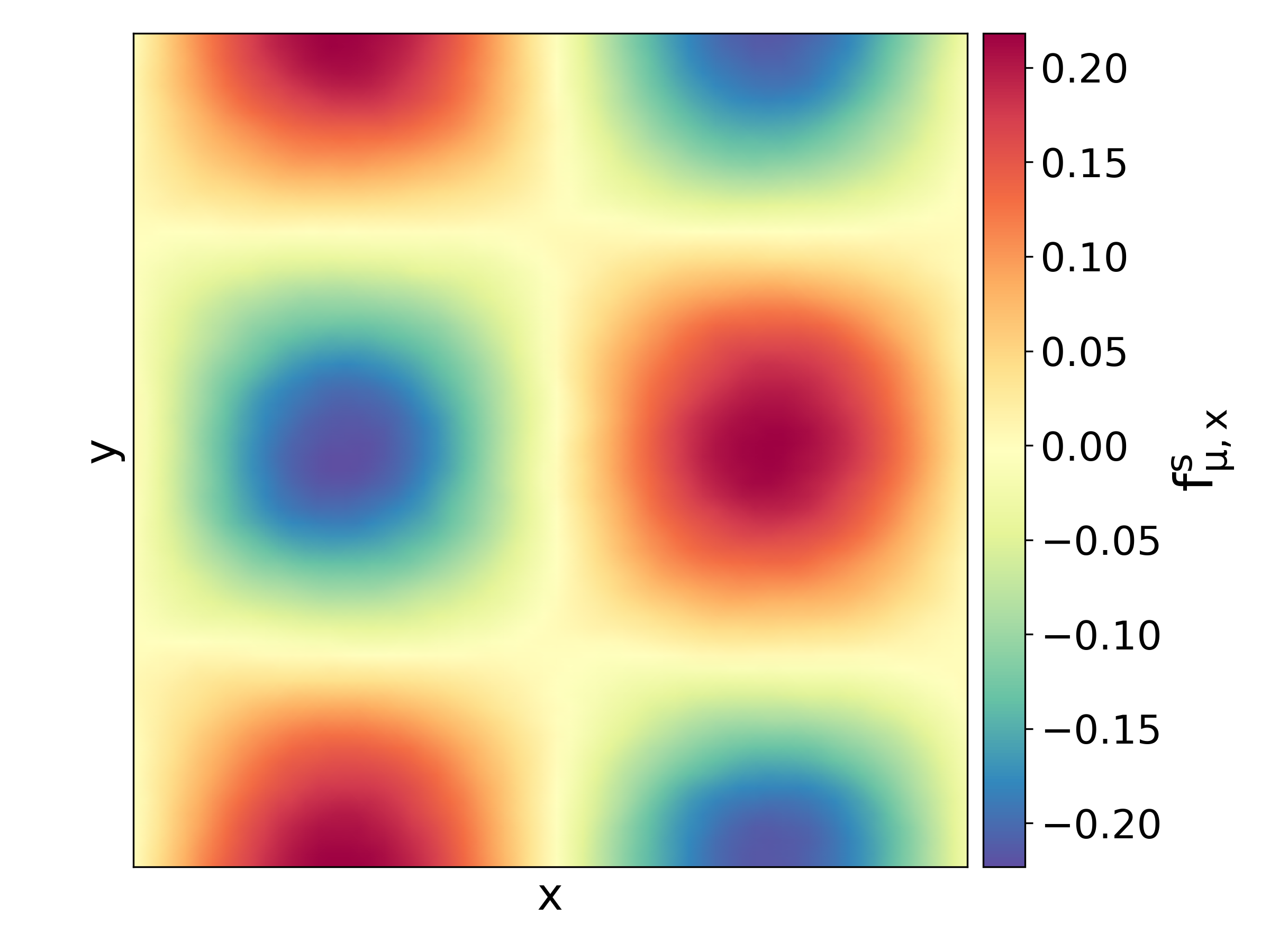}
        \caption{$f_{\mu,x}^s$}
        \label{fig:Fx_perp_adj_noise_Wi03}
    \end{subfigure}
    \hfill 
        \begin{subfigure}[b]{0.49\textwidth}
        \centering
        \includegraphics[width=\textwidth]{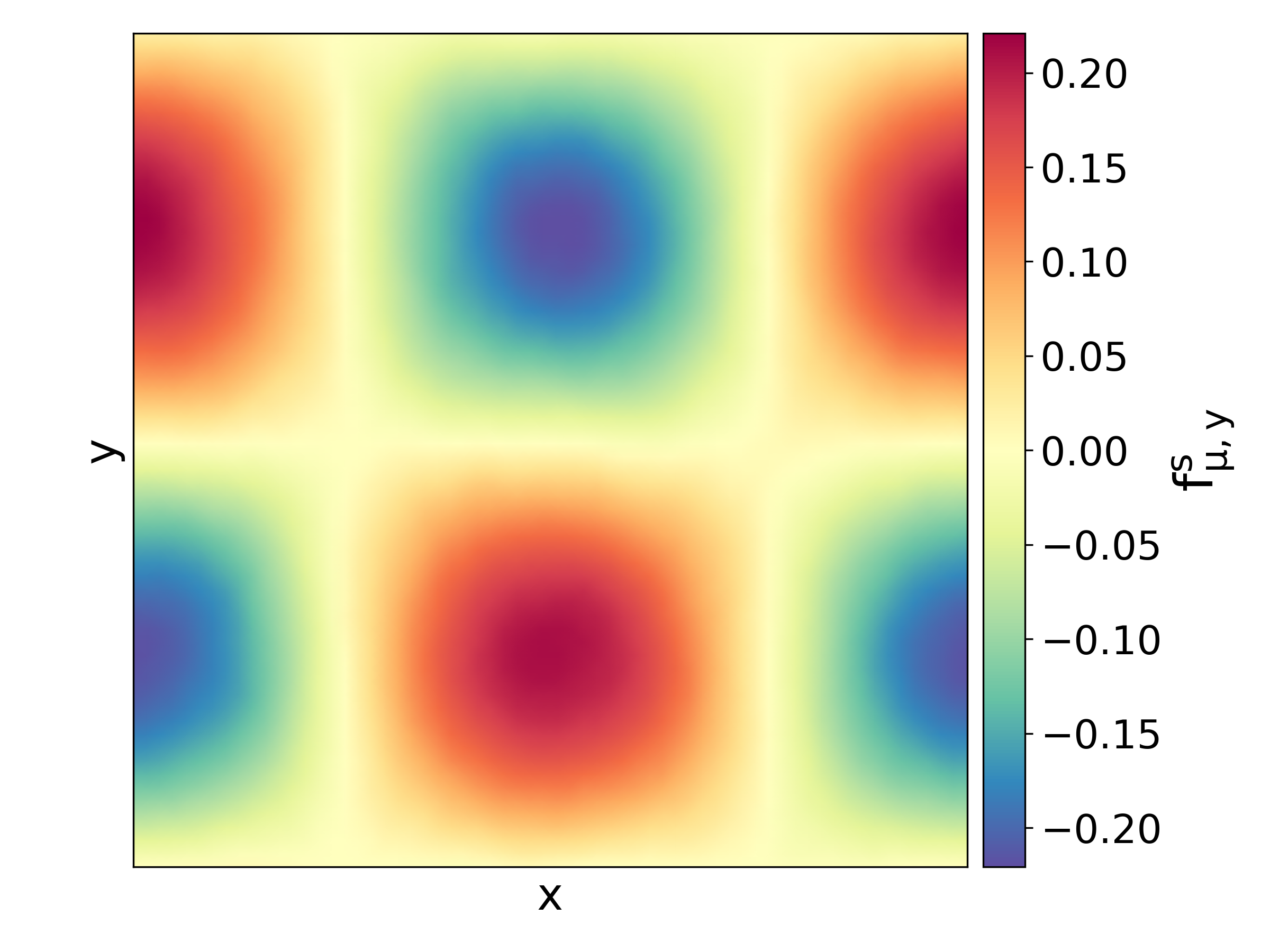}
        \caption{$f_{\mu,y}^s$}
        \label{fig:Fy_perp_adj_noise_Wi03}
    \end{subfigure}

    \vspace{0.5cm} 

        \begin{subfigure}[b]{0.49\textwidth}
        \centering
        \includegraphics[width=\textwidth]{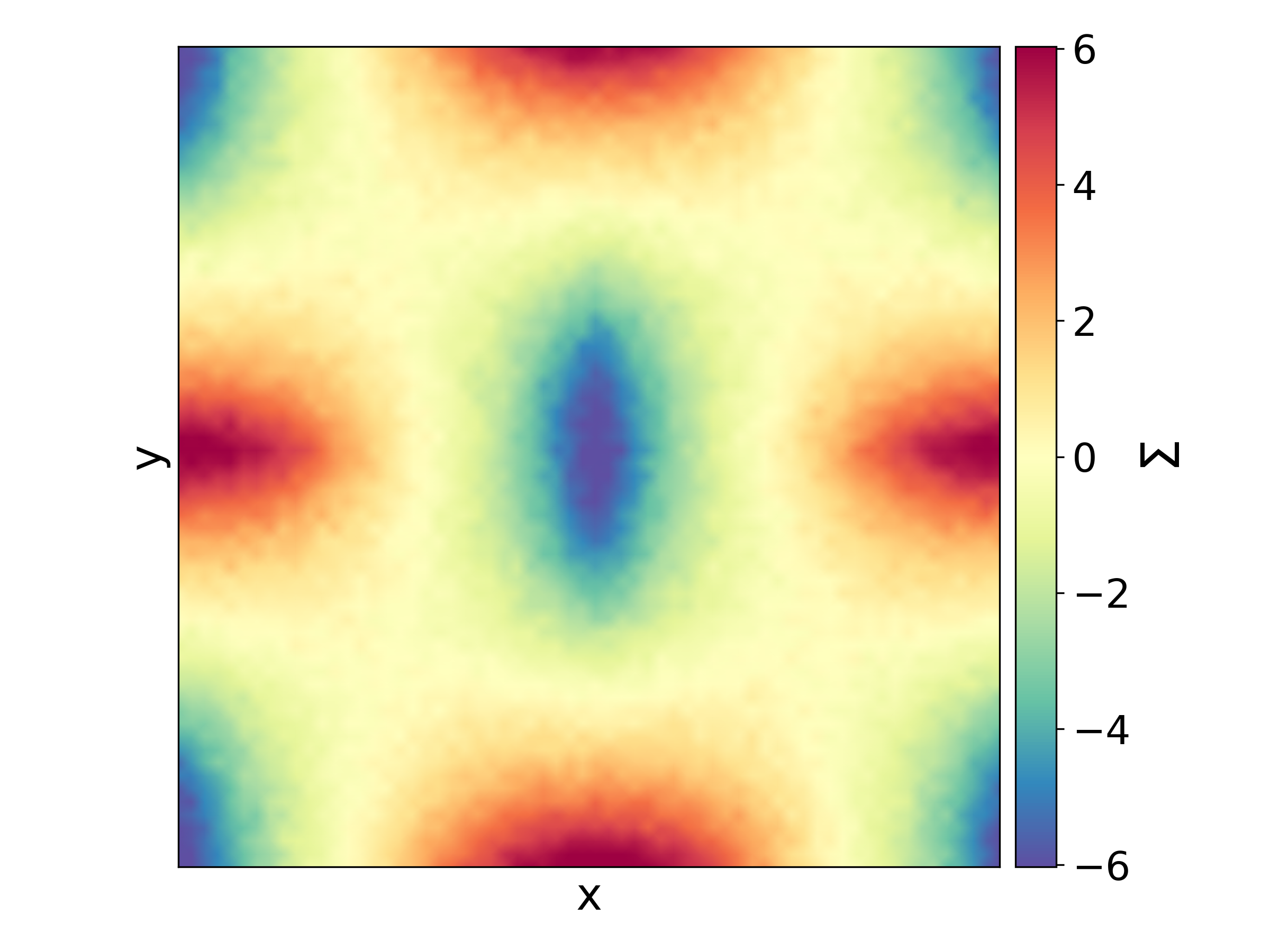}
        \label{fig:Fx_perp_strive_noise_Wi03}
        \caption{Reconstructed $\Sigma$}
    \end{subfigure}
    \hfill 
        \begin{subfigure}[b]{0.49\textwidth}
        \centering
        \includegraphics[width=\textwidth]{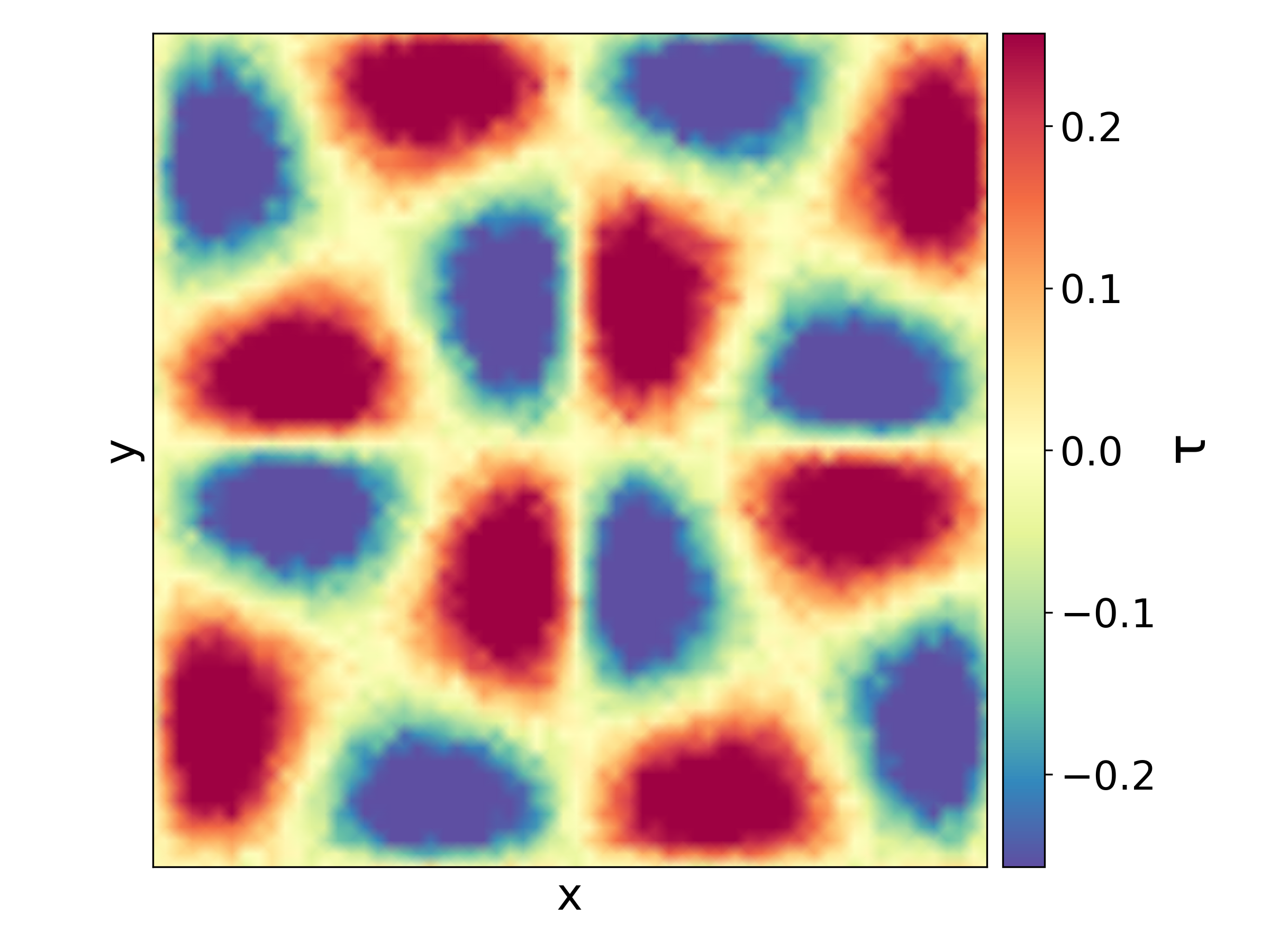}
        \label{fig:Fy_perp_strive_noise_Wi03}
        \caption{Reconstructed $\tau$}
    \end{subfigure}
        
    \caption{(a,b) Predicted force density from a noisy velocity field and gradient-regularized data assimilation. (c,d)  STRIVES-predicted stresses from a noisy microstructure field $\balpha$. }
    \label{fig:noisyfields}
\end{figure}

While these results illustrate the performance of STRIVES with regard to $\Sigma$ and $\tau$, experiments of course do not give access to these.   Rather, the data assimilation method of Sec.~\ref{sec:DAforForceDensity} yields $\fmus$. As noted above, we can reconstruct an estimate $\hat{\bff}_\mu^s$ from the STRIVES result. Fig.~\ref{fig:fmudiff} shows this for the $x$-component of $\fmus$, as well as the pointwise error, which is $\sim1\%$; the result from data assimilation is shown in Fig.~\ref{fig:DAresults1}. Because of the symmetry of this system, results for the $y$-component are identical modulo a rotation of $\pi/2$. 

The above results were obtained in the absence of noise. Noise in the velocity measurements can be mitigated by addition of a gradient regularization, Eq.~\eqref{eq:fgradient}. Fig.~\ref{fig:noisyfields}a and b shows the predicted $\fmus$ with $\lambda_n= 10^{-6}$, when the velocity field is polluted at each point with zero-mean Gaussian noise with standard deviation $\sigma = 0.05$.  Noise in the microstructure requires no modification to the STRIVES formalism for determining $\tL$, since it is already posed as a regression problem.  Adding zero-mean Gaussian noise with standard deviation $\sigma = 0.1$ to each component of $\balpha$ leads to the STRIVES predictions for $\Sigma$ and $\tau$
shown in Fig.~\ref{fig:noisyfields}c and d; these compare well to the noise-free results of Fig.~\ref{fig:stressreconWi03}. Note that in none of the results shown in Fig.~\ref{fig:noisyfields} have we performed any filtering on the actual data. Even without doing so, the STRIVES formalism yields accurate results, indicating its robustness to noise.

\section*{Conclusions}
This work has shown that the stress in a general 2D flow of a complex fluid can be revealed from velocity and microstructure data supplemented with physical and mathematical information.  Only when all of these elements are combined can the stress be determined; without any one of them, it would remain hidden. To clearly elucidate the method and provide proof-of-principle, here a relatively simple, idealized example has been used. Future work will generalize STRIVES to a finite element setting applicable to arbitrary planar geometries, as well as to more general stress-microstructure relationships, using well-developed methods for nonlinear regression such as neural networks or Gaussian processes. These generalizations will enable application of STRIVES to a wide range of complex flows of complex fluids.  Furthermore, the STRIVES framework can be naturally extended to incorporate multimodal data including for example, rheometry, rheo-optic or rheo-scattering measurements, and pressure measurements. As a specific application, for fluids with a direct, but perhaps complex, relationship between optical anisotropy and stress, STRIVES can use birefringence to learn this relationship and infer the stress without assuming a linear stress-optic rule.

Therefore, the possibilities for both fundamental and applied impact are enormous, as STRIVES enables an entirely new experimental capability: ``stress imaging.'' With respect to the fundamentals of non-Newtonian fluid mechanics, STRIVES enables observation of stress in poorly understood complex systems such as those undergoing yielding, flow-induced phase transitions, and shear-banding \cite{Barrat20241010882515-7639ad06cc}. It provides an 
opportunity to test models for stress-microstructure-deformation relationships in complex flows rather than just viscometric ones. In engineering applications, STRIVES, along with the data-driven SIMPLE formalism for data-driven microstructural evolution equations \cite{Young2023101007s00397-023-01412-0} represents a major step toward digital twins for flow process evaluation and design of complex fluids. This technology will impact understanding and design in important contexts such as porous medium flows found in applications from geological resource extraction to subcutaneous drug delivery, as well as coating processes for high-performance applications such as flexible electronics and solar cells.

\section*{Acknowledgments}
MDG acknowledges support from the National Science Foundation (CBET-2347344, CBET-2437151, DMR-2309000 (UW MRSEC)). MEH acknowledges support from the U.S. Department of Energy, Office of Science, Basic Energy Sciences under Award No. DE-SC0021294.  We thank Jake Buzhardt for helpful comments.

\bibliography{references,MDGLibrary,MDGnewbib}

\end{document}